\documentclass{article}
\usepackage{ltwol2e}
\usepackage{amsmath}
\usepackage{amssymb}
\usepackage{epsfig}

\def\be{\begin{equation}}
\def\ee{\end{equation}}
\def\bea{\begin{eqnarray}}
\def\eea{\end{eqnarray}}

\begin{document}

\title{
New Results on the Theoretical Precision of the LEP/SLC
Luminosity\thanks{Work supported in part by the US DoE, contract 
DE-FG05-91ER40627, and by the Polish Government, grant KBN 2P30225206.}
}

\author{ B.F.L.\ Ward,
S.\ Jadach\thanks{Permanent address: Institute of Nuclear Physics,
          ul. Kawiory 26a, PL 30-059 Cracow, Poland},
          M. Melles\thanks{Current address: Department of Theoretical Physics,
          University of Durham, South Road, Durham City DH1 3LE, England}
     ,  and S.\ A.\ Yost\\}

\address{{\em Department of Physics and Astronomy, 
          University of Tennessee,}\\
      {\em Knoxville, Tennessee 37996-1200, USA}}

\author{UTHEP-98-0501\\
       Sept., 1998}

\address{Invited talk presented by B.F.L. Ward at the 1998 Rochester Conference, Vancouver, July, 1998}


\twocolumn[\maketitle\abstracts{ We consider the error budget for the calculation of the
LEP/SLC luminosity in the Monte Carlo event generator BHLUMI4.04
from the standpoint of new calculations of the exact result for the
${\cal O}(\alpha)$ correction to the process $e^+ e^-
\rightarrow e^+ e^- + \gamma$ in the low angle luminosity regime at
SLC/LEP energies, for the double bremsstrahlung effect 
$e^+ e^- \rightarrow e^+ e^- + \gamma\gamma$ in this regime, and
for the size of the two-loop virtual correction to
$e^+ e^- \rightarrow e^+ e^-$ in this regime in context of
Yennie-Frautschi-Suura exponentiation. We find that the error
on the ${\cal O}(\alpha^2)$ photonic correction can be reduced from
the currently published value $0.1\%$ to the value $.027\%$.
This leads to an over-all precision tag for the currently available
program BHLUMI4.04 of $0.061\%$. This reduction of the precision
of the calculation is important for the final LEP1 EW precision
Z physics tests of the Standard Model.}] 


Currently, new luminometers at LEP\cite{lumin:1996} have made
measurements of the luminosity process $e^+ e^- \rightarrow e^+ e^- +
n(\gamma)$ at the experimental precision tags below $.1\%$.  This
should be compared with the prediction by the Knoxville-Krakow (KK)
Collaboration in the program BHLUMI4.04~\cite{bhl4:1996} wherein the
theoretical precision tag of $0.11\%$ is realized for this process
in the ALEPH SICAL-type~\cite{sical} acceptance. If one combines the
experimental results, one arrives at an experimental precision
of $\lesssim 0.05\%$. Evidently, for the final EW precision tests
data analysis for LEP1, it would be desirable to reduce the
theoretical precision tag on the luminosity cross section prediction
at least to the comparable $.05\%$-regime in order 
not to obscure unnecessarily
the comparison between experiment and the respective Standard Model
of the electroweak interaction. With this as our primary motivation,
we have examined the error budget arrived at in 
Refs.~\cite{bhl4:1996,arbuzov:1996,lepybk96} in view of recent exact results
impacting both the technical and physical precision of the errors
quoted in that budget.\par

More precisely, if one looks into the error budget shown in Table 1
Ref.~\cite{bhl4:1996,arbuzov:1996}, one sees that the largest contribution
is associated with the ${\cal O}(\alpha^2)$ photonic corrections,
which contribute $0.1\%$ in quadrature to the total $0.11\%$ quoted
for the total precision of the BHLUMI 4.04 prediction in these
references for the ALEPH SICAL-type acceptance. Accordingly, we have 
used the exact results in Refs.~\cite{1r1v:1996,2brem:1992,2brem:1993} and the
exact result in Ref.~\cite{frits:1988} to make a more realistic
estimate of the true size of this dominant error quoted in
Refs.~\cite{bhl4:1996,arbuzov:1996}.\par

In re-examining the photonic corrections used in BHLUMI 4.04
at the ${\cal O}(\alpha^2)$, which is the relevant order of the
corrections, one needs look at the approximations made in the
matrix element used in the calculation encoded in the program
in comparison to available exact results. This will allow us
to re-assess the physical precision of the corresponding
part of the BHLUMI 4.04 matrix element, which is the exact
${\cal O}(\alpha^2)$ LL(leading-log) Yennie-Frautschi-Suura (YFS)
exponentiated matrix element. The implementation of the 
Monte Carlo algorithm in BHLUMI 4.04 for two hard photon emission
needs also to be checked at this level of precision,
since our previous checks on it do not cover sufficiently the
two hard photon phase space as we were always working in the
leading-log approximation for two hard photons. 
This check, which we have recently completed, will 
allow us to give a more realistic estimate of the
technical precision of the realization of the corresponding aspect of the
matrix element in BHLUMI 4.04. The net result is a new estimate
of the total precision of the prediction of the luminosity cross section
by BHLUMI 4.04 at LEP1 energies.\par 

Our discussion is organized as follows. We first discuss the effect of
including the exact result in Ref.~\cite{1r1v:1996}
for the ${\cal O}(\alpha)$ correction
to the single hard bremsstrahlung process in BHLUMI4.xx in comparison
to the LL result for this correction that is used in BHLUMI4.04.
We then turn the the technical precision test of the implementation
of the two hard bremsstrahlung matrix element in BHLUMI4.04, wherein
this matrix element is also computed in the LL approximation. We will
carry-out this technical precision test in comparison to the analogous
test when the exact two hard bremsstrahlung matrix 
element~\cite{2brem:1992,2brem:1993} is implemented
in BHLUMI4.xx. This will verify that indeed the physical precision of
the LL approximation for the two hard bremsstrahlung matrix element
is indeed small in comparison to the other errors in the error budget
in Table 1 of Ref.~\cite{bhl4:1996,1r1v:1996}. Finally, we turn to the
effect of including the exact two-loop virtual correction in BHLUMI4.xx
in comparison to the LL approximation of the ${\cal O}(\alpha^2)$
virtual correction that is used in BHLUMI4.04. By combining the
results of these analyses, we arrive at a more realistic estimate
of the error on the theoretical prediction for the luminosity
process at LEP1/LEP2 energies as it is calculated by BHLUMI4.04.
\par

Considering now the exact ${\cal O}(\alpha)$ correction
to the single hard bremsstrahlung in the luminosity process,
we have implemented the results in Ref.~\cite{1r1v:1996} into BHLUMI4.xx
and made a systematic study of the net change in the prediction
for the luminosity relative to the prediction of BHLUMI4.04
in which this correction is treated to the LL level. What we find is
illustrated in Figs. 1-2 for the ALEPH SICAL-type acceptance
at the $Z^0$ peak. In the language of the YFS theory, this correction
enters the hard photon residuals as $\bar\beta^{(2)}_1$, the 
${\cal O}(\alpha^2)$ contribution to the one-hard photon residual
$\bar\beta_1$. In the Fig.~1, we show this part of the SICAL-type
accepted cross section as it is given by our 
exact result in Ref~\cite{1r1v:1996}
and as it is given by several different approximations to
our exact result: the LL approximation in BHLUMI4.04,
the approximate ansatz in Eq.(3.25) of Ref.~\cite{1r1v:1996},
and the result (NLLB) of Ref.~\cite{EAK:1996} which is 
calculated using a semi-collinear approximation that
the respective authors of Ref.~\cite{EAK:1996} argue 
includes the LL and NLL effects.
\begin{figure}
\center
\epsfig{file=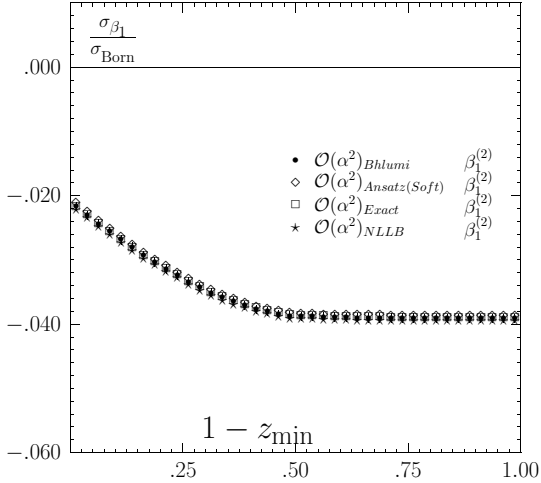,height=2.5in}
\caption{\small\sf 
Monte Carlo result ($10^6$ events) for the entire cross 
section associated with
$\bar\beta^{(2)}_1$ for the SICAL Wide-Narrow trigger.The first and second
order results are divided by the Narrow-Narrow Born cross section.
$z_{min}$ is as it is defined in Fig.~2 of Phys. Lett. {\bf B353}(1995)362.
}
\label{fig:beta1-1}
\end{figure}

In Fig.~2, we show the
difference between the corresponding LL result in BHLUMI4.04
and the other three results in Fig.~1 in ratio to the respective Born
cross section
.
\begin{figure}
\center
\epsfig{file=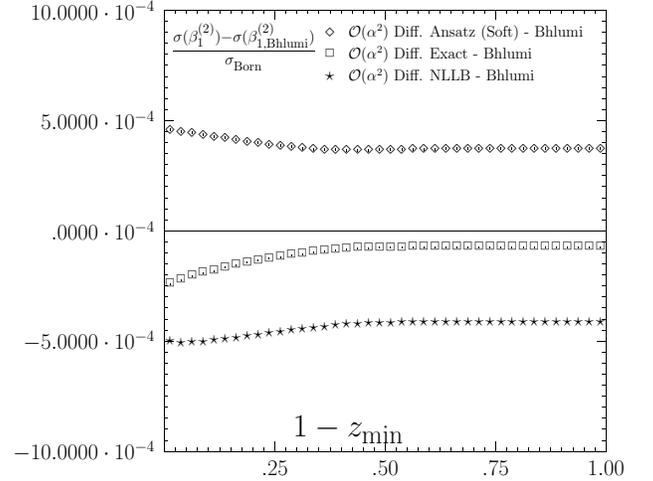,height=2.5in}
\caption{\small\sf
Pure second order Monte Carlo result for $\bar\beta^{(2)}_1-
\bar\beta^{(2)}_{1,Bhlumi}$ differences for the SICAL Wide-Narrow
trigger, divided by the Narrow-Narrow Born cross section,
with $z_{min}$ as given in Fig.~1.
}
\label{fig:beta1-2}
\end{figure}%
%
What we see is that the BHLUMI4.04 results are within $.02\%$ 
of the exact result in units of the respective Born cross section
throughout the experimentally interesting regime $0.2\le 1-z_{min}\le 1.0$.
This is the main reason we will be able to reduce the estimated precision
of the BHLUMI4.04 prediction in comparison to 
Ref.~\cite{bhl4:1996,arbuzov:1996}.\par

Turning next to the technical precision of the 2-$\gamma$ 
bremsstrahlung calculation in BHLUMI4.04, we have constructed a completely
independent realization of the two photon phase space integration 
compared to what is used in BHLUMI4.04 by way of an independent 
Monte Carlo algorithm. We have implemented this new Monte Carlo
realization of the two photon phase space and compared its result
with that of BHLUMI4.04's for the hard photon residual $\bar\beta_2$
contribution to the luminosity cross section, both for the LL
matrix element in BHLUMI4.04 and for the exact matrix element
for the two-photon bremsstrahlung in Ref.~\cite{2brem:1992,2brem:1993}.
We stress that the two photon phase space in BHLUMI4.04 is exact.
What we find is shown in Fig.~3 for the ALEPH
SICAL-type acceptance at the $Z^0$-peak.
\begin{figure}
\center
\epsfig{file=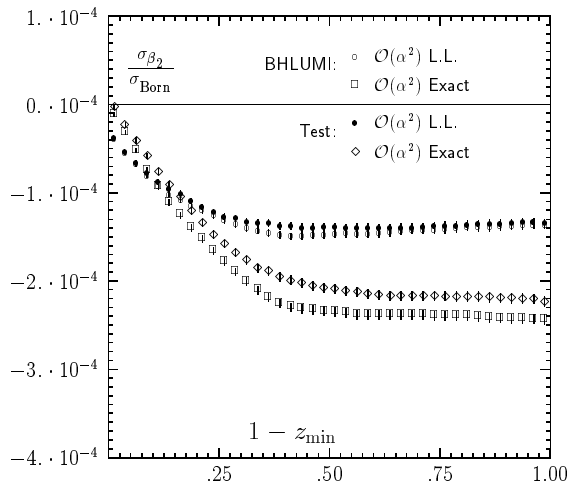,height=2.5in}
\caption{\small\sf
Comparison of Monte Carlo results for $\bar\beta^{(2)}_2$
for the LL and exact matrix
elements for $10^6$ events. Results are shown for the BHLUMI generator
and for an alrenative 'Test' generator for a technical precision test.
The results are for the SICAL Wide-Narrow trigger, and are divided by the
Narrow-Narrow Born cross section; $z_{min}$ is as given in Fig.~1.
}
\label{fig:process}
\end{figure}
%
We find that the
difference between the two realizations of the 2-$\gamma$ bremsstrahlung
is below $0.003\%$ of the Born cross section. Moreover, we get
an estimate of the physical precision of the LL approximation for
this part of the cross section from comparing the LL and exact results
as $0.012\%$, in agreement with our estimate in Ref.~\cite{bhl4:1996}.\par

Finally, we turn to the exact result for the two-loop contribution
of the hard photon residual $\bar\beta_0$ to the cross section in comparison
to the LL result used for it in BHLUMI4.04.  We have analytically continued
the result of Ref~\cite{frits:1988} from the s-channel to the
t-channel for the required two-loop contribution to the respective
charge form factor in QED. In this way, using the YFS theory, we have
found that the difference between the LL result in BHLUMI4.04
and the exact result corresponds to the shift of the function
$\upsilon$ in Eq.(2) in Ref.~\cite{acta:1996} by
\begin{align}
\Delta\upsilon^{(2)}& = ({\alpha\over\pi})^2L\left(6+6\zeta(3)-{45\over8}
                        -{\pi^2\over 2}\right) \notag\\
                    & + ({\alpha\over\pi})^2[6-9\zeta(3)+({17\over8}-2\ln 2)\pi^2
                        -{8\over45}\pi^4]\notag\\
\end{align}
where the big logarithm is defined as $L=\ln |t|/m_e^2$ and $\zeta(3)$
is the Riemann $zeta$ function of argument $3$. For the ALEPH SICAL type
acceptance at the $Z^0$ peak, this corresponds to $0.014\%$ in the 
cross section.\par
    Collecting the results above in quadrature, we obtain the result
that the current calculation of the ${\cal O}(\alpha^2)$ photonic 
corrections in BHLUMI4.04 are accurate to
\begin{equation}
{\Delta\sigma_{\cal L}\over \sigma_{\cal L}}|_{{\cal O}(\alpha^2)-photonic}=
    0.027\%  .
\end{equation}
Using this result in Table 1 for Ref.~\cite{bhl4:1996} we arrive at
the precision tag  $0.061\%$ for the currently available calculation
in BHLUMI4.04 at the $Z^0$ peak. At the LEP2 energy of $176$GeV, if we
repeat the analysis just described, we find that the corresponding
precision of BHLUMI4.04, for both the SICAL and LCAL type acceptances,
is now reduced to $0.122\%$ compared to the estimate in Ref.~\cite{bhl4:1996}
of $0.25\%$. The current situation is now summarized in our Table 1.

\begin{table*}[hbtp]
\centering
\begin{tabular}{|l|l|l|l|l|}
\hline
 & \multicolumn{2}{|c|}{LEP1} & \multicolumn{2}{|c|}{LEP2} \\
\hline
  Type of correction/error
& Past~\protect\cite{lepybk96,arbuzov:1996}
& Present
& Past~\protect\cite{lepybk96,arbuzov:1996}
& Present     \\
\hline
(a) Missing photonic 
    ${\cal O}(\alpha^2 )$~\protect\cite{elsewh} &
    0.10\%      & 0.027\%    & 0.20\%  & 0.04\%
\\
(b) Missing photonic 
    ${\cal O}(\alpha^3 L^3)$~\protect\cite{th-96-156} &
    0.015\%     & 0.015\%    & 0.03\%  & 0.03\%
\\
(c) Vacuum polarization~\protect\cite{burkhardt-pietrzyk:1995,eidelman-jegerlehner:1995} &
    0.04\%      & 0.04\%    & 0.10\%  & 0.10\%
\\
(d) Light pairs~\protect\cite{pairs:1993,th-96-244} &
    0.03\%      & 0.03\%    & 0.05\%  & 0.05\%
\\
(e) Z-exchange~\protect\cite{th-95-74}   &
    0.015\%      & 0.015\%   &  0.0\%  & 0.0\%
\\
\hline
    Total  &
    0.11\%      & 0.061\%    & 0.25\%  & 0.122\%
\\
\hline
\end{tabular}
\caption{\small\sf
Summary of the total (physical+technical) theoretical uncertainty
for a typical
calorimetric detector.
For LEP1, the above estimate is valid for the angular range
within   $1^{\circ}-3^{\circ}$, and
for  LEP2  it covers energies up to 176~GeV, and
angular range within $1^{\circ}-3^{\circ}$ and $3^{\circ}-6^{\circ}$
.
}
\label{tab:total-error-lep}
\end{table*}

A more detailed exposition of the results in this paper will appear
elsewhere~\cite{elsewh}.
\par
Our result that the size of the error associated with the 
missing sub-leading bremsstrahlung
correction at ${\cal O}(\alpha^2)$ in BHLUMI4.04 is $.027\%$ agrees
with the estimate of $0.03\%$ made by Montagna {\it et al.}~\cite{oreste} using
a structure function convolution of a hard collinear external
photon with an acollinear internal photon. As these authors have argued,
while such a pairing of convolutions does not represent a complete set
of photonic ${\cal O}(\alpha^2L)$ corrections, one expects it to
contain the bulk of such corrections. Indeed, our exact result of
$0.027\%$ shows that indeed the approximation made in Ref.~\cite{oreste}
does give the bulk of the respective ${\cal O}(\alpha^2L)$ correction.
Evidently, that we now have two independent results, one exact, that
presented by us in this paper,
and one approximate, that in Ref.~\cite{oreste}, which agree on the
size of the error associated with the missing photonic 
${\cal O}(\alpha^2L)$ correction in BHLUMI4.04 enhances the 
results in this paper.\par 

\section*{Acknowledgments}
Two of the authors (S.J.\ and B.F.L.W.) would like to thank 
Prof.\ A.\ De Rujula of the CERN TH Div.\ and 
the ALEPH, DELPHI, L3  and OPAL  
Collaborations, respectively, for their support and 
hospitality while this work was completed. 
B.F.L.W.\ would like to thank Prof.\ C.\ Prescott of Group A at SLAC for his 
kind hospitality while this work was in its developmental stages.\\
\vskip0.5cm
\noindent
$^a$~Work supported in part by the US DoE, contract 
DE-FG05-91ER40627, by the Polish Government, grant KBN 2P30225206
and by the USA-Poland MCS Joint Fund II, No. PAA/DOE-97-316.\\
$^b$~Permanent address: Institute of Nuclear Physics,
          ul. Kawiory 26a, PL 30-059 Cracow, Poland\\
$^c$~Current address: Department of Theoretical Physics,
          University of Durham, South Road, Durham City DH1 3LE, England\\


\section*{References}
\bibliographystyle{unsrt}

\end{document}